\documentstyle[preprint,aps,epsf]{revtex}
\begin{document}
\title{Interface charged impurity scattering in semiconductor MOSFETs and 
MODFETs: temperature dependent resistivity and 2D ``metallic'' behavior}

\author {S. Das Sarma, E. H. Hwang, and Igor \v{Z}uti\'{c}}
\address{Department of Physics, University of Maryland, College Park,
Maryland  20742-4111 } 
\maketitle

\begin{abstract}
We present the results on the anomalous 2D transport behavior by 
employing Drude-Boltzmann transport theory and taking 
into account the realistic charge impurity scattering effects.
Our results show quantitative agreement with the existing experimental 
data in several different systems and address the origin of the strong
and non-monotonic temperature dependent resistivity.

\keywords{Metal-insulator semiconductor structures, 
Metal-insulator transition}


\end{abstract}

\section{Introduction}

A large number of recent experimental publications on low temperature 
transport measurements in low-density high mobility two dimensional (2D)
electron systems 
in Si MOSFETs\cite{one}, GaAs MODFETs\cite{two}, and SiGe 
hetrostructures\cite{three} report an 
anomalously strong temperature dependent resistivity in the narrow regime of 
$0.1 - 5K$. In contrast to the usual Bloch-Gr\"{u}neisen theory 
of essentially a temperature-independent low-temperature resistivity,
the measured resistivity changes by as much as 
a factor of ten for a $1-2K$ increase in temperature.
This observed anomaly has led to a great deal of theoretical activity 
\cite{four,five}
involving  claims of an exotic metal or even a superconducting 
system at the interface producing the strong temperature dependent 
resistivity, which has no known analog in ordinary three dimensional 
metallic behavior. Much more interest has focused around the possibility 
of a 2D metal-insulator quantum phase transition being responsible 
for the observed strong temperature dependent resistivity since
theoretically a 2D electron system at T=0 has so far been thought to be 
(at least in the absence of electron interaction effects) an insulator 
\cite{six}.

In this paper 
we provide a theoretical explanation for the temperature dependent 
resistivity of the 2D systems
in the ``metallic'' phase ($n_s \ge n_c$, where $n_s$ is the 2D density and 
$n_c$ the critical density which separates ``metallic'' and ``insulating'' 
behavior) in the absence of magnetic field \cite{seven}
by using the Drude-Boltzmann transport theory with
RPA screening and the Dingle temperature 
approximation to incorporate collisional broadening effects 
on screening \cite{nine}.
In our approach we leave out quantum corrections, including 
localization effects, and neglect the inelastic electron-electron 
interaction, which may well be significant in the low density 2D 
systems of experimental relevance. 
Our calculated resistivity agrees quantitatively with
the existing experimental data \cite{one,two,three} on the temperature 
dependent low-density resistivity of 2D electron systems. We find that 
the strong temperature dependence arises from a combination of two 
effects: the strong temperature dependence of finite wave vector 
screening in 2D systems and a sharp quantum-classical crossover 
due to the low Fermi temperature in the relevant 2D systems. 

\section{Theory}

We use the finite temperature Drude-Boltzmann theory to calculate 
the ohmic resistivity of the inversion layer electrons, taking 
into account only long range scattering by the static
charged impurity centers with the 
screened electron-impurity Coulomb interaction.
The screening effect is included within the random 
phase approximation (RPA) with 
the finite temperature static RPA dielectric (screening) function 
$\kappa(q,T)$  given by
\begin{equation}
\kappa(q,T) = 1 + \frac{2\pi e^2}{\bar{\kappa}q}F(q)\Pi(q,T),
\end{equation}
where $F(q)$ is the form factor for electron-electron interactions and 
$\Pi(q,T)$ is the static polarization.
We assume that the charged impurity centers
are randomly distributed  in the plane parallel
to the semiconductor-insulator surface.
Within the Born approximation 
the scattering time $\tau(\varepsilon,T)$ for our model is given by
\begin{equation}
\frac{1}{\tau(\varepsilon,T)} = \frac{2\pi}{\hbar}\int\frac{d^2k'}{(2\pi)^2}
 \int^{\infty}_{-\infty}N_{i}(z) dz 
\left |\frac{v_q(z)}{\kappa(q,T)}\right |^2 (1-\cos\theta) \delta\left (
\epsilon_{\bf k} - \epsilon_{\bf k'} \right ),
\end{equation}
where $q = |{\bf k} - {\bf k}'|$, 
$N_{i}(z)$ is the impurity density of the charged center,
$\theta \equiv \theta_{{\bf kk}'}$
is the scattering angle between ${\bf k}$ and ${\bf k}'$, 
$\varepsilon = \epsilon_{\bf k} = \hbar^2k^2/2m$, 
$\epsilon_{\bf k'} = \hbar^2k'^2/2m$, $v_q(z)$ is the 2D electron-impurity
Coulomb interaction.
In calculating the Coulomb interaction and the RPA dielectric function 
in Eq. (1) we take into account subband quantization effects in the 
inversion layer through the lowest subband variational wavefunction.
The resistivity is given by $\rho = 
m/(ne^2 \langle \tau \rangle) $, where 
$m$ is the carrier effective mass, $n$  the effective free carrier density
\cite{five}, and $\langle \tau \rangle $  the energy
averaged scattering time. The average is given by 
$\langle \tau \rangle ={\int d\varepsilon \tau(\varepsilon) \varepsilon
\left ( -\frac{\partial f}{\partial \varepsilon}
\right )}/{\int d\varepsilon \left ( - \frac{\partial f}{\partial 
\varepsilon} \right )}\varepsilon$,
where $f(\varepsilon)$ is the Fermi distribution function, $f(\varepsilon) 
=\{ 1+\exp[ (\varepsilon-\mu)]/k_B T \}^{-1}$ with finite temperature chemical
potential, $\mu=\mu(T,n)$, which is determined self-consistently. 

\section{Results and Conclusion}

It is physically instructive to first consider the asymptotic behavior 
of the temperature dependent part of resistivity, $\rho(T)$.
In the quantum regime at low temperature, $T \ll T_F$ with $T_F \equiv 
\mu(T=0)/k_B$, the dominant behavior of $\rho(T)$ is linearly increasing
with T, i.e., $\rho(T) \propto T/T_F$ arising from the temperature 
dependent screening, $\kappa(q,T)$ \cite{nine}.
In the high temperature limit $(T \gg T_F)$ corresponding to the 
classical regime, the resistivity is decreasing with temperature, i.e.,
$\rho(T) \propto T_F/T$ due to the 
energy averaging of $\tau$.
For intermediate temperatures ($T \sim T_F$)
the system crosses over from a non-degenerate classical to a  
strongly screened degenerate quantum regime \cite{five}.

In Fig. 1 we give our numerically calculated resistivity $\rho(T,n)$ 
for the Si-12 sample of Ref. \cite{one} using
the effective carrier density $n=n_s-n_c$ \cite{five} at
several values of $n_s > n_c$ and different Dingle temperatures.
The impurity density, $N_i$, sets 
the overall scale of resistivity ($\rho \propto N_i$), and does not 
affect the calculated $T$ and $n$ dependence of $\rho(T,n)$. 
We obtain, at low densities, both the 
observed non-monotonicity  and  the 
strong drop in $\rho(T)$  in the $0.1 \sim 2 K$ 
temperature range\cite{one,two,three,eight}.  Our high 
density results show weak monotonically increasing $\rho(T)$ with 
increasing $T$ similar to experimental 
observations \cite{one,two,three}.
In the inset we show the analytic zero temperature conductivity 
as a function of density
$n_s$, following the approach of Ref. \cite{ten}. An approximately 
linear dependence is in a good agreement with the $T \rightarrow 0 $ 
extrapolation of the experimental \cite{one} resistivity.
Obtained results 
suggest that the reduced effective density and not the total value 
contributes to conductivity  and 
supports our basic freeze-out or binding model \cite{eleven}.
These analytic results  
coincide with the full numerical calculation, 
further justifying validity of our methods. 

\begin{figure}
\epsfysize=4.65in
\centering
\epsffile{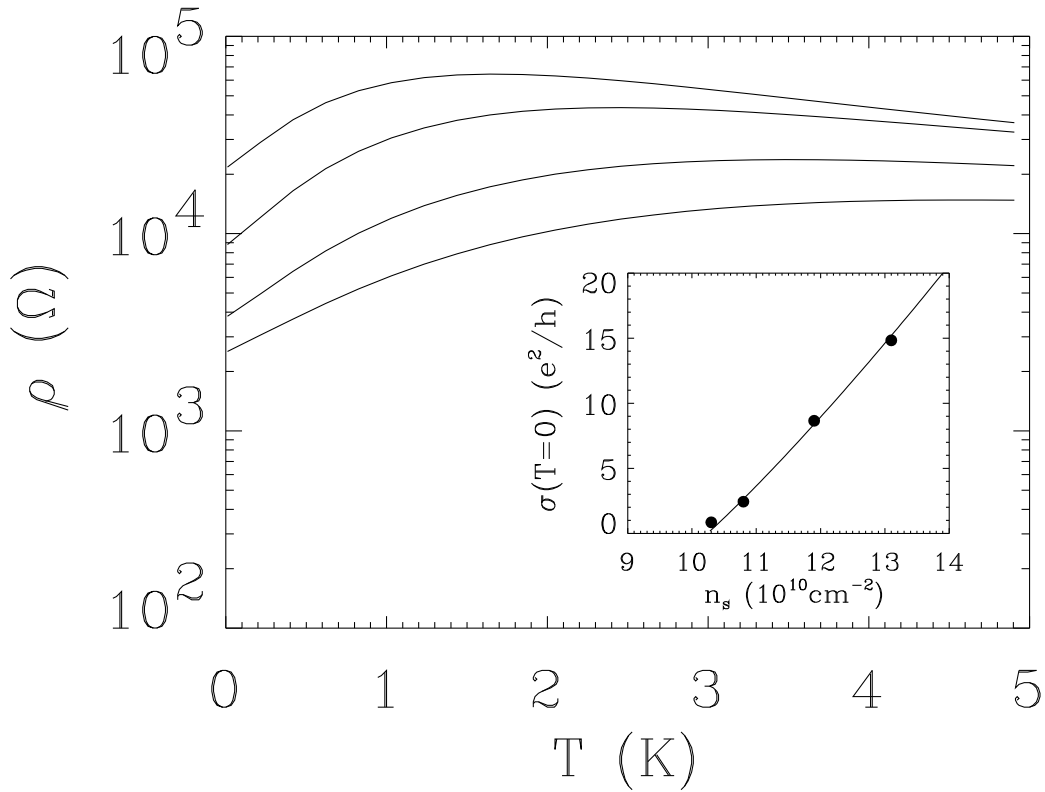}
\caption{
The calculated resistivities for various electron densities, 
$n_s = $ 1.03, 1.08, 1.19, 1.31 $\times 10^{11}cm^{-2}$ (top to bottom)
as a function of $T$ for the Si-12
sample of Ref. [1], using the critical density $n_c=10^{11}cm^{-2}$.
In the inset we show the analytic zero temperature 
conductivity as a function of density
$n_s$. Points represent extrapolated $\sigma(T \rightarrow 0)$  
from Si-12 sample of Ref. [1]. }
\end{figure}

In conclusion we obtain good agreement with the experimental 
results. The strong temperature dependence of resistivity at low 
and intermediate densities ($n_s \geq n_c$)
arises from the temperature dependent
screening and a low Fermi temperature by virtue of the low effective 
carrier density.
Thus, charged impurity scattering, carrier binding 
and freeze-out, temperature and density dependence of 2D screening, 
and classical to quantum crossover  are playing 
significant roles in the experiments and can not be neglected in
theoretical analysis of the ``2D M-I-T'' phenomenon.


{\it Acknowledgments} -- This work is supported by the U.S.- ARO 
and the U.S.- ONR.


\begin{thebibliography}{99}

\bibitem{one} 
S. V. Kravchenko {\it et al}., Phys. Rev. B {\bf 50}, 8039 (1994);
{\bf 51}, 7038 (1995); Phys. Rev. Lett {\bf 77}, 4938 (1996); 
D. Simonian {\it et al}., {\it ibid} 
{\bf 79}, 2304 (1997); V. M. Pudalov, JETP Lett. {\bf 65}, 932 
(1997); D. Popovic {\it et al}., Phys. Rev. Lett. {\bf 79}, 1543 (1997).

\bibitem{two} Y. Hanein {\it et al}., Phys. Rev. Lett. {\bf 80}, 1288 
(1998); M. Y. Simmons {\it et al}., {\it ibid} {\bf 80}, 1292 (1998).

\bibitem{three} P. T. Coleridge {\it et al}., Phys. Rev. B {\bf 56}, 
R 12764(1997); J. Lam {\it et al}., {\it ibid} {\bf 56}, R 12741 (1997).

\bibitem{four}V. M. Pudalov, JETP Lett. {\bf 66}, 175 (1997);
V. Dobrosavljevi\'{c} {\it et al}., Phys. Rev. Lett. {\bf 79},
455 (1997); C. Castellani {\it et al}., Phys. Rev. B {\bf 57}, 9381 
(1998); S. He and X. C. Xie, Phys. Rev. Lett. {\bf 80}, 3324 (1998); 
D. Belitz 
and T. R. Kirkpatrick, Phys. Rev. B {\bf 58}, 8214 (1998); Q. Si 
and C. M. Varma, Phys. Rev. Lett. {\bf 81}, 4951 (1998);
P. Phillips {\it et el}., Nature, {\bf 395}, 253 (1998); 
B. Altshuler and D. Maslov, Phys. Rev. Lett. {\bf 82}, 145 (1999).

\bibitem{five} S. Das Sarma and E. H. Hwang, Phys. Rev. Lett. 
{\bf 83}, 164 (1999); T. M. Klapwijk and S. Das Sarma, Solid. State Commun. 
{\bf 110}, 581 (1999); S. Das Sarma, E. H. Hwang, 
and I. \v{Z}uti\'{c}, in preparation.

\bibitem{six} E. Abrahams {\it et al}., Phys. Rev. Lett. {\bf 42}, 
673 (1979).

\bibitem{seven} S. Das Sarma and E. H. Hwang, cond-mat/9909452,
explains the parallel magnetic field effects 
on the quasi-2D transport behavior.



\bibitem{nine} T. Ando {\it et al}., Rev. Mod. Phys. 
{\bf 54}, 437 (1982); F. Stern, Phys. Rev. Lett. {\bf 44}, 1469 (1980); 
F. Stern and S. Das Sarma, Solid State Electron. {\bf 28}, 158 (1985).



\bibitem{eight} A. P. Mills {\it et al}., Phys. Rev. Lett. {\bf 83}, 2805
(1999).


\bibitem{ten}A. Gold and V. T. Dolgopolov, Phys. Rev. 
B {\bf 33}, 1076 (1986).

\bibitem{eleven}V. T. Dolgopolov {\it et al}., Phys. Rev. B {\bf 55}, R 7339 
(1997); Y. Hanein {\it et al}., {\it ibid}, {\bf 58}, R 7520 (1998). See also
S. Das Sarma and E. H. Hwang, cond-mat/9901117.

\end{thebibliography}
\end{document}